\documentclass{mem}
\usepackage{natbib}\usepackage{txfonts}\usepackage{balance}
\usepackage{graphicx}
\usepackage[a4paper]{hyperref}
\idline{75}{282}

\begin{document}
\def\teff{$T\rm_{eff }$}
\def\logg{$\log g$}
\def\kms{$\mathrm {km s}^{-1}$}

\title{
White Dwarf Spectra and Atmosphere Models
}

   \subtitle{}

\author{D. \,Koester\inst{1} }

  \offprints{D. Koester}

\institute{
Institut f\"ur Theoretische Physik und Astrophysik, Universit\"at
Kiel, 24098  Kiel, Germany
\email{koester@astrophysik.uni-kiel.de}
}

\authorrunning{Koester}

\titlerunning{White Dwarf Spectra}

\abstract{
We describe the spectral classification of white dwarfs and some
of the physical processes important for their understanding. In the
major part of this paper we discuss the input physics and
computational methods for one of the most widely used stellar
atmosphere codes for white dwarfs.
}
\maketitle{}

\section{Spectral classification of white dwarfs}
The classification scheme for white dwarfs (WD) developed in the
beginning in analogy to the main sequence spectral types, with a
distinguishing letter ``D'' for degenerate object. DAs thus were stars
with very strong Balmer lines, DBs had strong He\,I lines, DOs
He\,II. Today we know that this classification -- in contrast to the
main sequence -- has not much to do with effective temperature, but is
an indication of the photospheric composition. The classification used
today was developed and described in detail in
\citet{Sion.Greenstein.ea83}.

The main characteristic is the division into hydrogen-rich (DA)
and helium-rich (DB, DO) atmospheres, but again in contrast to normal
stars the most abundant element dominates with very few exceptions by
several orders of magnitude. The explanation for this quasi
mono-elemental composition is gravitational separation
\citep{Schatzman47}. In the absence of significant competing
macroscopic motions (stellar wind, meridional circulation, convection)
the heavier elements diffuse downward, leaving the lightest element
present floating on top. The helium-dominated objects in this scenario
must have lost their thin outer hydrogen envelope during the formation
phase of the white dwarf in the late stages of the asymptotic giant
branch or planetary nebular phase.

Besides the major types mentioned above one distinguishes DC (too cool
to show any spectral feature, mostly helium-rich), DQ (atomic or
molecular features of carbon), DAZ, DBZ, DZ (objects with traces of
metals in hydrogen-rich or helium-rich atmospheres). 

The carbon in the DQ is assumed to be dredged up from deeper layers by
the growing convection zone in the superficial helium layer
\citep{Koester.Weidemann.ea82, Pelletier.Fontaine.ea86}, whereas the
other heavy metals must be accreted from an outside source, either the
interstellar matter, or some debris from a tidally disrupted asteroid.

A spectral atlas showing many example spectra for all major types has
been published by \citet{Wesemael.Greenstein.ea93}.

\section{Observational quantities}
Stellar parameters (effective temperature \teff, surface gravity
\logg, abundances) are obtained from an analysis of spectroscopic or
photometric observations. If the surface of a star could be resolved,
as for the sun, and if all relevant properties of our instrument were
known, we could in principle determine the energy emitted by a small
element of surface area, per unit time, wavelength interval, solid
angle, into our line of sight. This quantity is called the (specific)
intensity, and in the case of a spherically symmetric star the only
geometric variable for the surface value is the angle of emission
relative to the normal on the surface element $\vartheta$, that is
\begin{equation}
I = I(\vartheta) \mbox{\qquad or \qquad} I = I(\mu)
\end{equation}
with $\mu =\cos \vartheta$. If we cannot resolve the surface we can
only measure the average intensity of the stellar disk $\bar{I}$. More
specifically, the energy flux $f$ arriving outside the terrestrial
atmosphere is related to this average intensity by

\begin{equation}
 f = \bar{I} \,\Omega
\end{equation}
with average intensity
\begin{eqnarray}
\bar{I}& = & 2\int\limits^{\pi/2}_0 I(\vartheta) \cos(\vartheta)
\sin(\vartheta)\, d\vartheta \nonumber \\
& = & 2\int\limits^1_0 I(\mu) \mu\, d\mu
\end{eqnarray}
and the solid angle of the star
\begin{equation}
\Omega = \frac{\pi R^2}{D^2}
\end{equation}
with radius $R$ and distance $D$.

If we want to determine stellar parameters from a comparison of
observed and theoretically calculated spectra, the quantity which has
to be calculated is thus the intensity $I$ at the surface of the
star. The theory of stellar atmospheres has been developed by many
authors over the past century and has reached a very mature state
today. Classical works, still worth reading, are
e.g. \citet{Unsold68} and \citet{Mihalas78}. ``Model atmospheres''
and ``synthetic spectra'' as well as computer codes to calculate them
are widely available. In the remainder of this paper we will describe
in detail the input physics and computational methods used by the
author for his model atmospheres, which are used by many groups. 

The programming of the code was started by Dr. Thomas Gehren about
1975 with minor contributions by myself. However, since then
practically every routine has been completely rewritten several times
by the current author, and every remaining programming error is only
my fault.

\section{Model atmospheres and synthetic spectra}
The basic procedure is to specify the element abundances in the
atmospheres, and the parameters effective temperature \teff\ and
surface gravity \logg, which are used as proxies for the ``typical''
values of the thermodynamic variables temperature and pressure. Using
a number of simplifying assumptions and basic laws of physics this is
sufficient to predict the radiation field (intensity) at the surface
of the star. The most important assumptions are

\begin{itemize}
\item {\bf homogenous, plane parallel layers:} the depth of the
  atmosphere is considered to be very small compared to the radius of
  the star. All matter quantities (density, pressure, temperature)
  depend only on one geometric variable, the height (in radial
  direction) $z$. The intensity depends on $z$ and the angle against
  the normal $\vartheta$, but not on the azimuthal angle.
\item {\bf hydrostatic equilibrium:} at each point within the outer
  layers, which have a direct influence on the emerging radiation
  (i.e. the atmosphere or photosphere) the gradient of the gas
  pressure is in equilibrium with the gravitational attraction (plus
  possibly the transfer of momentum by photons).
\item {\bf radiative and convective equilibrium:} there is no energy
  generation or loss within the atmosphere, only transport of the
  energy generated in the deep interior. This transport can occur
  through radiation, heat conduction, or convection; the total energy
  flux as determined by the parameter effective temperature is
  constant at all depths. 
\item {\bf Local Thermodynamic Equilibrium:} the matter is in thermal
  equilibrium corresponding to the local temperature at each layer,
  that is the ionization, excitation, dissociation of molecules
  etc. are governed by the usual relations of thermal equilibrium
  (Boltzmann factors, Saha equation, Kirchhoff's law etc.).  This is a
  very important assumption since it decreases the computational
  effort by several orders of \goodbreak magnitude. Please note that
  thermal equilibrium (i.e. the Planck function) is {\em not} assumed
  for the radiation field! Except for white dwarfs hotter than about
  50000~K this LTE assumption is well justified.
\end{itemize} 

The code is divided in two major parts, which calculate in turn the
physical structure of the outer layers (run of temperature, density,
pressure, absorption coefficients etc. with depth, this part will be
called ATM here) and the surface intensities for many wavelengths
(emerging spectrum, called SYN). There are auxiliary programs for
additional necessary tasks, e.g. one for calculating the equation of
state and absorption coefficients (KAPPA), another one for calculating
equivalent widths of spectral lines or theoretical magnitudes in any
photometric system (FILT), and so on.

\subsection{Equation of State (KAPPA)}
If no molecule formation has to be considered (e.g. high temperatures)
and no elements besides H and He are present, the thermodynamic
calculations are made directly in parallel with the determination of
the atmospheric structure in the program ATM. Otherwise, these
calculations are made in KAPPA and the results (tables of matter
density $\rho$, electron pressure $P_e$, entropy, absorption
coefficients etc.) are stored in large two-dimensional tables as
function of temperature $T$ and gas pressure $P_g$. The EOS, Saha
equation for ionization, and dissociation equilibria for molecules are
derived from a model Free Energy, which includes the ideal gas terms,
Coulomb corrections and an ``Excluded Volume'' term for the non-ideal
interaction of neutral particles. Electron degeneracy is tested in all
layers, but currently not implemented in the EOS, as it has been
unimportant in the range of parameters, where I have used my codes.

Partition functions for H, HeI, and HeII are explicitly calculated
using the lowest 100 levels from the TOPBASE database
\citep{Cunto.Mendoza92, Cunto.Mendoza.ea93} and applying the
occupation probability $w$ according to the prescriptions of
\citet{Hummer.Mihalas88, Mihalas.Dappen.ea88}. For all other elements
we use tables given by \citet{Kurucz70} providing the partition
function for a nominal cutoff 0.1 eV below the ionization limit.  The
actual limit is calculated by a hydrogenic fit to the higher levels
using a cutoff determined from the non-ideal terms in the EOS.

Currently dissociation equilibria are implemented for 20 molecules
(H$_2$, CH, NH, OH, MgH, SiH, CaH, C$_2$, CN, CO, N$_2$, NO, O$_2$,
  TiO, H$_2$O, HCN, HCO, C$_3$, CO$_2$, N$_2$O) using data from
  \citet{Kurucz70} and  \citet{Tatum66}. 

The non-linear system of Saha and dissociation equations together with
the condition of neutrality and the definition of total gas pressure
is solved with a Newton-Raphson iteration.

\subsection{Absorption coefficients (KAPPA)}
The absorption coefficient $\kappa$ describes the probability $w$ that
a photon will interact (be absorbed or scattered), when traveling a
small distance $ds$ in matter of density $\rho$
\begin{equation}
      w = \kappa \rho ds = \sum n_i\, a_i\, ds
\end{equation}
For a dimensionless $w$, $\kappa$ has to have the dimension of area
per mass. It usually is the sum of many different interaction
processes, with each contribution determined from the number density
of particles in the absorbing atomic state $n_i$ and the area $a_i$,
the cross section for this interaction. The most important processes
for white dwarfs and some sources of data or routines are (note that
in most cases the data have been transformed by us and/or new routines
written for our use):

\noindent
{\bf bound-free and free-free absorption of neutral hydrogen:} this
can be calculated quasi-classically and corrected to quantum mechanics
by the so-called Gaunt factors \citep{Menzel.Pekeris35,
Karzas.Latter61, Kurucz70}. The free-free coefficient for
all other ions (with the exceptions noted below) is calculated
hydrogen-like.

\noindent
{\bf bound-free and free-free absorption of the H$^-$ ion:} numerical
fits from \citet{John88}.

\noindent
{\bf bound-free and free-free transitions of the H$_2^+$ ion:} 
    data and numerical fits from \citet{Boggess59}.

\noindent
{\bf bound-free absorption of neutral helium:} 
cross sections for the 43 lowest levels of HeI are taken from the
TOPBASE database. 

\noindent
{\bf free-free absorption of He$^-$:} helium does not have a bound
state as negative ion, so only the free-free process is needed. Data
are from \citet{John94}.

\noindent
{\bf bound-free and free-free absorption of the negative carbon ion
  C$^-$:} data for the bf cross sections are from
\citet{Robinson.Geltman67} and \cite{Cooper.Martin62}, for the ff
cross section from \citet{John.Williams76}.

\noindent
{\bf bound-free transitions for elements other than H, He:} these
cross sections are mostly from the TOPBASE database if they are
available there, or hydrogen-like calculations otherwise.

\noindent
{\bf Thomson scattering by free electrons:} the constant cross section
per electron
\begin{equation}  \sigma = 6.6527\times 10^-{25} \mbox{\quad cm}^2
\end{equation} 
is used.

\noindent
{\bf Rayleigh scattering by HI, HeI, H$_2$:} cross section fits are from
\citet{Dalgarno62, Dalgarno.Williams62, Kurucz70}.

\noindent
{\bf Molecular absorption:} the calculations use the
just-overlapping-line or smeared-line approximation in the version
developed by \cite{Zeidler.Koester82}. This assumes that the density
and broadening of rotational lines are so high that they form a
quasi-continuum. Currently implemented are molecular data for C$_2$,
C$_3$, and H$_2$ molecules.

\noindent
{\bf Spectral line absorption:} atomic data (excitation energies,
oscillator strengths, line broadening constants) are obtained from a
number of atomic databases, predominantly the line lists from Kurucz
and collaborators \citep{Kurucz.Bell95}, and the VALD (Vienna)
database \citep{Kupka.Ryabchikova.ea00, Kupka.Piskunov.ea99,
  Ryabchikova.Piskunov.ea97, Piskunov.Kupka.ea95}.

Satisfactory theories and data for the line profiles do exist for the
Stark broadening of neutral hydrogen \citep{Lemke97,
  Vidal.Cooper.ea73}, and for 21 optical lines of neutral helium
  \citep{Barnard.Cooper.ea69, Beauchamp.Wesemael.ea97}. These are
  so-called ``unified theories'', which attempt to describe the total
  line profile from core to the far wing. Similarly, the first three
  Lyman lines of H broadened by ionized and neutral perturbers and
  including a number of satellite features are well described by the
  work of Nicole Allard and collaborators \citep[e.g.][and many
    earlier papers]{Allard.Kielkopf.ea04}.

For all other processes the situation is much less satisfactory. Stark
broadening parameters for further HeI lines are provided by
\citet{Dimitrijevic.Sahal-Brechot90}. In many later papers of the Belgrade
group around Dimitrijevic similar data are provided for other
elements. 

Below 8000~K for hydrogen-rich and 16000~K for helium-rich atmospheres
line broadening by neutral particles becomes important. Since in most
objects we have one dominating element, the interaction is usually
between H-H or He-He. Resonance broadening is thus important
\citep{Ali.Griem65}, as well as van der Waals interaction. Only the
first three Balmer lines \citep{Barklem.Piskunov.ea00} and some He
transitions \citep{Leo.Peach.ea95} have so-called self-broadening
theories; in the latter case however for very low temperatures (300~K)
only. These theories combine the effects of resonance and van der
Waals broadening in a more consistent way.

A few experimental measurements of broadening constants do exist, but
in the vast majority of metal lines the Stark and van der Waals
broadening constants can only be very roughly estimated by simple
approximations \citep[e.g.][]{Unsold68, Cowley71, Griem66}. The line
profile in the ``impact approximation'' is then described by a Lorentz
profile with these damping constants, which yields a Voigt profile
after convolution with a Doppler profile for the line-of-sight
velocities of the emitting atoms.

\subsection{Atmospheric structure}
As mentioned above, the basic quantity for the description of the
radiation field is the intensity
\begin{equation}
                   I = I(z,\lambda,\mu)
\end{equation}
with the geometrical height scale $z$ measured from an arbitrary level
outward of the star, wavelength $\lambda$, and cosine of the angle
against the z-axis $\mu = \cos \vartheta$. Useful quantities derived
from this are the mean intensity (averaged over all directions, not to
be confused with the disk-averaged intensity $\bar{I}$)
\begin{equation} 
                  J = \frac{1}{4\pi} \oint I\,d\omega
                    = \frac{1}{2} \int\limits_{-1}^1 I\,d\mu
\end{equation}
and the energy flux by radiation per unit area
\begin{equation}
                   F = 2\pi \int\limits_{-1}^1 I \mu\, d\mu
\end{equation}
Another useful quantity is
\begin{equation}
                   K = \frac{1}{2} \int\limits_{-1}^1 I \mu^2 d\mu .
\end{equation}
Note, that at the surface of a spherically symmetric star with no
radiation from the outside we have
\begin{equation}
                   F = \pi \bar{I}
\end{equation}
that is, the energy flux through the surface of a star is the quantity
to be calculated for the comparison with non-resolved observations of
a white dwarf.

\subsubsection{The equation of radiative transfer}
\begin{figure*}[t!]
\resizebox{\hsize}{!}{\includegraphics[clip=true]{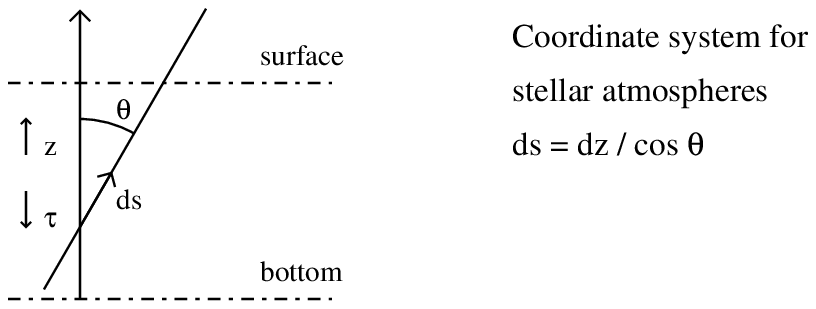}}
\caption{\footnotesize
Geometry for the radiative transfer equation}
\label{geo}
\end{figure*}
The equation of radiative transfer describes the balance between
emission and absorption of photons along the path $ds$, using the
geometry described in Fig.~\ref{geo}
\begin{eqnarray}
      dI &=& \rho \varepsilon ds - I \rho \kappa ds \nonumber \\ 
      \mu \frac{dI}{\rho \kappa dz} &=& \frac{\varepsilon}{\kappa} - I 
       \nonumber \\
      \mu \frac{dI}{d\tau} &=& I - S.
\end{eqnarray}
$\varepsilon$ and $\kappa$ are the emission and absorption
coefficients per mass, $\tau$ is a new depth variable called optical
depth, replacing the geometric variable $z$, $d\tau = - \rho \kappa
ds$. $\tau = 0$ corresponds to the top of the atmosphere. $S$ is the
so-called source function, $ S = \varepsilon/\kappa$.  

For the solution of this differential equation two boundary conditions
have to be specified. For the incoming radiation at the top $I(0,\mu)
= 0 $ for $\mu < 0$ is usually assumed. At the bottom, at some large
value $\tau_B \gg 1$ the incoming intensity from below $I(\tau_B,\mu)$
for $ \mu>0$ has to be specified. One possibility is to assume that at
large depth the source function is the Planck function $B$ (see
below), expand it around the value $\tau_B$ and derive from the
transfer equation
\begin{equation}
I(\tau,\mu) = B(\tau) + \mu \frac{dB}{d\tau}
\end{equation}
or higher order approximations.

For the absorption and emission coefficients we have even at this
phenomenological level of description to distinguish between two
different processes. In the case of absorption they are called ``true
absorption'' $\kappa_t$ and ``scattering'' $\sigma$, the corresponding
emission processes are ``thermal emission'' $\epsilon_t$ and again
scattering $\epsilon_s$. Scattering means that a photon through
interaction with matter changes its direction, but not the energy,
whereas in the case of ``true absorption'' energy is absorbed by the
matter and possibly re-emitted later with a different energy;
scattering processes thus do not lead to an energy coupling of the
radiation field with matter.

In our case of LTE Kirchhoff's law states
\begin{equation}
\frac{\epsilon_t}{\kappa_t} = B
\end{equation}
an extremely powerful result. On the other hand, for isotropically
distributed scattering particles we can derive
\begin{equation}
 \epsilon_s = \sigma J
\end{equation}
If both kinds of processes are important we can write the source
function as
\begin{equation}
S = \frac{\varepsilon}{\kappa} = \frac{\epsilon_t +
  \epsilon_s}{\kappa_t + \sigma} = \frac{\kappa_t}{\kappa_t + \sigma}
B + \frac{\sigma}{\kappa_t + \sigma} J
\end{equation}
Sometimes only the case with $\sigma = 0$ and therefore $ S=B$ is
called LTE, or strict LTE. Since the inclusion of scattering in the
source function is not very difficult computationally, we do not make
this distinction here.

Because of the nature of the boundary conditions, given on both ends
of the solution interval, the solution of the first-order equation is
numerically difficult. Special precautions have to be taken to avoid
exponentially increasing parasitic solutions. This is very ingeniously
avoided by the method of \cite{Feautrier64}.
We introduce new variables by dividing $I$ into a symmetric ($u$) and
an antisymmetric ($v$) part (we use $\mu > 0$ and write $-\mu$ for the
negative angles)
\begin{eqnarray}
   u(\tau,\mu) & = & \frac{1}{2} \left[ I(\tau,\mu) + I(\tau,-\mu)\right] \\
   v(\tau,\mu) & = & \frac{1}{2} \left[ I(\tau,\mu) - I(\tau,-\mu)\right] .
\end{eqnarray}
It is clear from the definitions that $u(\tau,-\mu) = u(\tau,\mu)$ and
$v(\tau,-\mu) = -v(\tau,\mu)$, so we need the solution only for
positive $\mu$. Once $u$ and $v$ are known, we can always recover $I$
as well as $J$ and $F$. Writing the radiative transfer equation
separately for positive and negative $\mu$, adding and subtracting the
two, we can derive the Feautrier equations
\begin{eqnarray}
   \mu^2\,\frac{d^2u}{d\tau^2} &=& u - S \\
    v &=& \mu \frac{du}{d\tau}
\end{eqnarray}
The boundary conditions can easily be transformed to the new
variables. The numerical solution of the second order transfer
equation above is much easier and stable than for the first-order
equation.

\subsubsection{Further constraints}
The solution of the transfer equation needs the values of $B$
and absorption coefficients at each depth of the atmosphere, and
therefore the thermodynamic variables e.g. $T$ and $P_g$. The
additional constraints we have are the constant value of the
transported energy flux and the hydrostatic equation
\begin{equation}
F_{tot}(z) = \int \limits_0^\infty F(z, \lambda)\, d\lambda +
F_{conv}(z) = \sigma_R T^4\rm_{eff }
\end{equation}
Here $\sigma_R$ is the radiation constant of Stefan's law and
$F_{conv}$ the convective energy flux (see below).

The balance between the gradient of the gas pressure, gravitational
force and radiative force is
\begin{equation}
\frac{dP_g}{dz} = - \rho g + \frac{1}{c} \int\limits_0^\infty
\kappa(\lambda)\, F(\lambda)\, d\lambda
\end{equation}
where the second term on the right side describes the momentum
transferred from the radiation field to the matter. 
Defining a ``standard'' absorption coefficient $\kappa_S$ at a
standard wavelength, or as a weighted mean over wavelength (e.g. the
Rosseland mean), we can use the associated standard optical depth
scale $d\tau_S = - \rho \kappa_S dz$
\begin{equation}
\frac{dP_g}{d\tau_S} = \frac{g}{\kappa_S} - \frac{1}{c \rho \kappa_S}
\int\limits_0^\infty \kappa(\lambda)\, F(\lambda)\, d\lambda .
\end{equation}
The hydrostatic equation thus provides a relation between pressure
scale, geometrical, and optical depth. For technical reasons we use
the gas pressure as the independent variable, and the depths $z$ and
$\tau_S$ at each layer are derived quantities.

Since $\rho, F_{conv}$, and the absorption coefficients depend also on
temperature, the typical method of solution is to assume a temperature
stratification $T(P_g)$ and energy fluxes $F$ (e.g. from a previous
similar calculation or iteration step) and solve the two constraint
equation above together with the radiative transfer. These equations
together provide just enough equations for the unknowns, if the
temperature structure is known. Since this is in general not the case,
an iterative solution is necessary. The temperature dependent
quantities are expanded around the current value, e.g.
\begin{equation}
     B(z,T,\lambda) = B(z,T_0,\lambda) + \frac{dB}{dT}\,\Delta T
\end{equation}
The whole system of equations is then solved at once for the
temperature corrections $\Delta T$ and iterated with an improved
temperature stratification, until the corrections become sufficiently
small and all constraints are fulfilled. 

As can be seen from eqs.(21,22) the constraints couple all
wavelengths, which is responsible for the huge number of unknowns.  If
the knowledge of the detailed angle dependence of $I$ or $u$ is not
needed the computational burden in some intermediate steps can be
considerably reduced by the method of ``variable Eddington factors''
\citep{Auer.Mihalas70}.  We start from the Feautrier
equation eq.(19) and integrate over $\mu$ from 0 to 1
\begin{equation}  {{d^2K}\over{d\tau^2}} = J - S
\end{equation}
Under many conditions, in particular at large optical depths, the
ratio $K/J$ tends to a constant value $1/3$. We introduce a ``variable
Eddington factor'' $f = K/J$ to get
\begin{equation} {{d^2fJ}\over{d\tau^2}} = J - S .
\end{equation}
Assuming $f$ to be known, the structure of this equation is the same
as that of the original Feautrier equation and can be solved with the
same methods. The value of $f$ has of course to be calculated from the
original equation, but this can be done for one wavelength a time and
therefore much faster.

\subsubsection{Convection}
Convection under the conditions of white dwarfs is highly
turbulent. There is as yet no satisfactory theory describing the
energy transport from first principles, nor any realistic numerical
simulation,  which could be implemented in routine calculations of
atmospheric models. One has therefore to resort to the very crude
mixing-length approximation, originally by \citet{Prandtl25}, and
adapted to stellar conditions by \citet{Bohm-Vitense58}.

In the calculation of stellar evolution or even stellar atmospheres
for ``normal stars'' our colleagues are generally content with one
free parameter to describe the energy flux by convection in the
mixing-length approximation. This parameter is the ratio of the mixing
length to the pressure scale height $\alpha = l/H_P$. In the case of
white dwarfs we have gone further and use three numbers $a, b, c$,
which appear in the heuristic derivation of the theory, as additional
free parameters. Different versions of the MLT are then denoted as
e.g. ML1/${\alpha=1}$, or ML2/${\alpha=0.6}$, where ML1, ML2 describe
the choice of $a, b, c$, and $\alpha$ the mixing-length
\citep{Fontaine.Villeneuve.ea81, Tassoul.Fontaine.ea90,
  Jordan.Koester.ea98}. Comparison of UV with optical spectra of ZZ
Ceti DA white dwarfs around \teff\ = 11000 - 12000 ~K has shown that a
consistent description is possible with ML1/2.0
\citep{Koester.Allard.ea94} or ML2/0.6 \citep{Bergeron.Wesemael.ea95},
both of which describe what is called ``intermediate efficiency''
convection. The latter choice is at present used as quasi standard for
DAs, also by this author. It is, however, quite clear that MLT in
general is a very poor approximation and WD parameters are therefore
still uncertain, when convection is important.

\subsubsection{Numerical solution}
The solution is obtained by a discretization of the depth scale $P_g$,
the wavelengths $\lambda$, and angles $\mu$. Typical numbers for the
grid points are 4 values for $\mu$ between $0$ and $1$, 100 depth
points, and 1000 to 100000 wavelength points. Derivatives are
approximated by difference quotients, and integrals by sums. For the
integration over angles (to obtain $J, F$) Gaussian quadratures are
used for higher accuracy with few points. For the integration over
wavelength simple trapezoidal rule or Simpson's rule are used. We then
obtain a huge system of linear equations for the variables $I$ at each
depth, wavelength, and angle, and the $\Delta T$ at all
depths. 

Fortunately the matrix of this system is very sparse -- a
tridiagonal band structure of sub-matrixes and some extra lines and
columns from the constraint equations. \citet{Rybicki71} has
demonstrated a very efficient elimination scheme, which results in a
final linear system of rank equal to the number of depth points
(e.g. typically 100), which is full and has to be solved by standard
methods to determine the temperature corrections.  When these
corrections are deemed small enough (criteria used are often that the
relative temperature corrections are smaller than 0.001 and the total
flux at each depth is correct to 0.1 percent), the atmosphere
structure is determined and all important quantities (temperature, gas
pressure, electron pressure, density, specific heat, adiabatic
gradient, number densities of molecules, absorption coefficients) as
function of depths are saved in a file for further use.

\subsection{Synthetic spectra (SYN)}
The calculation of the atmospheric structure with ATM needs of course
also the radiation field, including the spectrum emerging from the
surface. The reasons why we use a separate program SYN to calculate
this again are the following:

For the calculation of the atmospheric structure all wavelengths are
coupled through the constraint equations, thus limiting the number to
typically a few 1000. On the other hand, because of the necessity to
calculate the total energy flux, the wavelength grid has to cover a
large range from X-ray to far infrared. For the comparison with
observations we typically need only a smaller range, but with much
higher wavelength resolution. As the structure is now known, we do not
need to consider the wavelength coupling again, but can calculate the
radiation field for each wavelength independently.

Because of this reduced burden we are free to use many more (even
weak) spectral lines, or more sophisticated line broadening
theories. We can also include much more detailed calculations of
molecular absorption bands.

\subsubsection{Numerical method}
The emerging spectral energy distribution $F(0,\lambda)$ could be
calculated using the Feautrier equations. However, for technical
reasons we use a different method here. Integrating the original
transfer equation over angle $\mu$ we can derive an integral equation
for the flux   
\begin{eqnarray}
F(\tau)  &=&  2 \pi \int\limits_\tau^\infty S(\tau')\, 
  E_2(\tau'-\tau)\, d\tau' \nonumber \\
   &-& 2 \pi \int\limits_0^\tau S(\tau')\, E_2(\tau-\tau')\, d\tau'  .
\end{eqnarray}
with the exponential integral function $E_2$.
In abbreviated form we write this as the flux integral operator $\Phi$
\begin{equation} 
F(\tau) = \Phi\left[S(\tau)\right] .
\end{equation}
A special case is the flux emerging from the surface of the star,
which as we know is equal to the observational quantity disk-averaged
intensity $F(0) =\pi \bar{I}$
\begin{equation}
F(0) = 2 \pi \int\limits_0^\infty S(\tau')\, E_2(\tau')\, d\tau'.
\end{equation}
In strict LTE $S(\tau) = B(T(\tau))$, which is known from the
atmospheric structure, and the calculation would be reduced to a
simple integration. In general, however, the source function may
include a scattering term and we need the mean intensity $J$. Formally
this can be derived directly from the transfer equation in a similar
way as for $F$, with the result
\begin{equation}
J(\tau) = {1\over 2} \int\limits_0^\infty S(\tau')\, E_1(|\tau -
     \tau'|)\, d\tau' 
\end{equation}
with the exponential integral $E_1$. The integral operator in this
equation is called the $\Lambda$ operator
\begin{equation}
 J(\tau) = \Lambda\left[S(\tau)\right] =  \Lambda\left[\alpha B +
   (1-\alpha) J \right].
\end{equation}
We call this a formal solution, since $S$ on the right hand side also
contains the unknown $J$, which makes this an integral equation.

For the numerical solution the depth scale is discretisized again,
transforming the continuous variables $J, B$ into vectors and the
$\Lambda$ operator into a matrix. We use an 18-point Gaussian
integration formula; the emerging flux $F(0,\lambda) = \pi \bar{I}$ as
well as the intensity $I(0,\lambda,\mu)$ can finally be calculated
from the source function $S$ by a simple integration (summation).

\subsection{Theoretical photometry and equivalent widths}
The final results of SYN are stored in a binary disk file. This file
contains a table of fluxes (or intensities) as a function of vacuum
wavelengths, the ``synthetic spectrum'' at the stellar surface, and in
addition the basic parameters and structure data of the atmosphere
model. These data are in a format, which can be used as input for the
ATM program to start the iteration for a similar model.

Auxiliary programs are available to transform these data, e.g. to air
wavelengths, or into an ASCII file for use by other authors. These
``export'' files contain the flux table and a header similar to the
FITS headers with all important parameters of the calculation. I
strongly encourage my users to never separate this header from the
table.

A program FILT calculates equivalent widths of spectral lines from the
flux table. It can also calculate theoretical photometry in arbitrary
filter systems as e.g.
\begin{equation}
      V = -2.5 \log \int \limits_0^\infty \bar{I}(\lambda)\,
      S_V(\lambda) \, d\lambda   + C_V
\end{equation}
with the total transmission $S_V$ of the filter plus optics,
terrestrial atmosphere, etc. The constant $C_V$ has to be determined
from standard stars with known absolutely calibrated spectrum
and  measured magnitude in the corresponding system. Very often Vega
is used for this purpose.
    
\section{Some very technical remarks and outlook}
The code is currently written in the programming language FORTRAN77,
but slowly -- as time permits -- transformed to FORTRAN95, which is
much less prone to programming errors and much easier to
maintain. Although considered a very old-fashioned programming
language by many (who most likely never used it), I have been able to
use my code over more than 30 years on dozens of computers and
operating systems, in most cases without ever changing a single line
of code. I am very grateful for that and do not plan to ever change to
another language.

Since more than twenty years, and through very fundamental changes,
the complete code has been under control of a version control system,
starting with SCCS, RCS, and about two years ago changed to
MERCURIAL. This means that I can recall the complete programs for any
date or any version back to about 1985, IF I know the relevant data
(version, compilation date, calculation date, etc.). These identifiers
are written into every output file, including the ASCII format of the
synthetic spectra, which are so widely distributed. 

These files also contain the information for some free parameters,
like the mixing-length version used, or changes to the Hummer-Mihalas
occupation probabilities. For this reason I urge users, to always keep
the header with the spectrum table, such that the code version used
can be identified in case of problems or questions. Unfortunately the
system is not perfect, since it does not keep track of a few data
files, which have to be changed for different calculations, the most
important being the file with the spectral line data. Different
databases provide quite often different data for oscillator strengths
or broadening constants. Depending on what I believe at the time of
calculation to be the most reliable values, these change from time to
time, and I cannot always reconstruct what has been used after some
years. I am thinking how to solve this problem, but so far without
result.

From the programming aspect, as mentioned above the code is slowly
moved to FORTRAN95. It is already very modular, with many modules free
of any side effects and being reused unchanged in different
programs. Programming the way I now know it should be was a hard
learning experience over decades, and I am glad that FORTRAN95 supports
almost all my ideas and preferences much better than the older
versions. The current aim is to make the whole program system very
user-friendly to be able to put it in the public domain under a GPL or
similar license in a few years.

\begin{acknowledgements}
I want to thank the organizers of the School of Astrophysics
"F. Lucchin" in Tarquinia (Italy) in June 2008 for inviting me to this
very pleasant experience. I am also grateful to get the chance to
describe my white dwarf atmosphere model calculations in a widely
available journal, for the information of the users. 

I am grateful to Dr. Thomas Gehren, who got me interested in stellar
atmospheres (though not for white dwarfs), and to the large number of
colleagues, who have used my calculations and contributed to many
improvements through their demands.
\end{acknowledgements}


\end{document}